\documentclass[a4paper,12pt]{article}
\usepackage{amsmath,amsfonts}
\usepackage[english]{babel}
\usepackage{a4wide}
\usepackage{url}
\usepackage{hyperref}
\usepackage{amssymb,amsthm}
\usepackage{booktabs}
\usepackage{graphicx}
\usepackage{threeparttable}
\usepackage{color}
\usepackage{longtable}
\usepackage{booktabs}
\usepackage{multirow}
\usepackage{bigdelim}
\usepackage{longtable}

\usepackage[cp1251]{inputenc}

\theoremstyle{plain}

\theoremstyle{definition}
\title{Statistical analysis of the remaining human life expectancy. Mathematical models of the type Gompertz-Makeham and others}

\author{Kateryna Akbash$^a$ (corresponding author), Ivan Matsak$^b$}

\date{}

\begin{document}
\maketitle

\begin{center}
$^a$ Kryvyi Rih State Pedagogical University, Ukraine \\

$^b$ Independent researcher, Ukraine\\
\medskip
kateryna.akbash@gmail.com, ORCID 0000-0003-3676-4574 (K.Akbash), i.m.k@ukr.net, ORCID 0000-0003-3816-7650 (I.Matsak)

%(e-mail:  ivanmatsak@univ.kiev.ua)
\end{center} \bigskip

\hspace{5,0cm}

\medskip

\bigskip

%\begin{quote}

\begin{center}
{Abstract}\\
\end{center}
%\vspace{0.5cm}
Many researchers believe that human lifetime on the interval (30,80) years is described by the Gompertz-Makeham distribution. In this work, the task is set to test this hypothesis. We consider a number of distributions: Gompertz-Makeham, Gompertz-Weibull, Logistic Kannisto-Makeham, Logistic Beard-Makeham and Logistic Beard-Weibull, some of which are partially known, for which comparative calculations on real mortality data are performed using the $\chi^2$ criterion. The data were selected for three countries: England \& Wales, Norway and Sweden, cohorts born in 1900. As it turned out, on the interval (30,80) years the Gompertz-Makeham distribution is far from being the best among those listed above. In addition, for the same data, a similar problem was studied for the interval (80,120) years.

%\end{quote}
\medskip

\bigskip
\noindent{\it Keywords:}  residual, lifetime,  statistical models

\medskip

\noindent{\it MSC:} {62P05,  60E05}

\bigskip

\maketitle

\section{Introduction. The main results.} \label{section1}
The question of life expectancy has always worried humanity. But systematic research into it began somewhere at the end of the 17th century and is associated with the presentation of data on human mortality in the form of life expectancy tables (J. Graunt, E. Halley). Based on these tables, the average survival time to a certain age was calculated.

It should be noted that such great scientists K. Huygens, 
W. Leibnitz, L. Euler, P. Laplace were interested in statistical data on mortality and their presentation in the form of life expectancy tables (see \cite{gav-91}, \cite{ros}).

Let us denote by $ \mathfrak{T}_{t_0}$ the remaining time of human life, which has reached a certain age  $t_{0}$. We will assume that $ \mathfrak{T}_{t_0}$ is a non-negative random variable (r.v.).

How is r.v. $ \mathfrak{T}_{t_0}$ distributed?

One of the first most successful attempts to mathematically describe the dependence of mortality on age belongs to the English actuary B. Gompertz (\cite{gom}, 1825). While studying real data, he noticed that human mortality is a logarithmic function of age.

But experience over the past several hundred years has shown that it's difficult to hope to accurately describe human life expectancy $ \mathfrak{T}_{t_0}$ using any known probability distribution. Therefore, both in the past and today, attempts have been made to find a simple distribution that at least approximately adequately describes real human mortality data over some fixed interval.

For example, a significant number of biologists and gerontologists believe that somewhere in the 30-80 year range, the Gompertz-Makeham law (see section \ref{section2} below) often describes real data such as human lifespan better than other well-known formulas (\cite{com}, \cite{gav-91}, \cite{ros}). However, renowned gerontologist B. Strehler \cite{str}, for example, suggests choosing an interval of 35-90 years, calling it the Gompertz period.

Unfortunately, we are unaware of any studies that have substantiated this hypothesis for the 30-80 year interval using rigorous statistical criteria. The formal theoretical basis for this assumption is as follows: Gompertz's law is closely related to the distribution of extreme values for minima (type 3 according to B.V. Gnedenko's classification \cite{gne43}, \cite{llr}). Furthermore, even the first calculations and empirical graphs of mortality rates show that the Gompertz-Makeham distribution approximates actual data on overall mortality significantly better than well-known distributions such as the normal, exponential, and Weibull distributions.

The question naturally arises: is there a simple distribution at all that describes data on the remaining life expectancy of a person in the interval (30, 80) years more accurately than the Gompertz-Makeham distribution?

In this article we will consider several simple distributions, mainly known as Gompertz-Makeham (GM), Gompertz-Weibull (GW), Logistic Kannisto-Makeham (LKM), Logistic Beard-Makeham (LBM) and Logistic Beard-Weibull (LBW), for which corresponding comparative calculations are carried out on real mortality data using the $\chi^2$ criterion.

Data for 3 countries will be considered: England \& Wales, Norway and Sweden. The authors tried to select reliable and homogeneous data whenever possible. As the results obtained  show (see sections \ref{section4}, \ref{section5}), over an interval of (30, 80) years, the distribution of GM is far from the best among them.

Another question that is logical to formulate is: what distribution adequately describes the remaining duration of human life after 80 years?

This problem was considered, for example, in the works \cite{tha1998}, \cite{am24}. In the book \cite{tha1998}, six possible mortality models were compared over the (80, 120)-year interval, which were fitted to a number of data sets. The data sets concern the periods 1960-1970, 1970-1980, and 1980-1990, as well as for cohorts born in 1871-1880, separately for men and women. They cover more than 32 million deaths at age 80 and older during this period. The authors argue that the data presented in \cite{tha1998} are clearly closer to the Logistic Beard-Makeham than to the Gompertz, Weibull, and Heligman models, and Pollard. An important contribution to the study of mortality at advanced ages is due to Thatcher \cite{tha}, where a model describing the increase of mortality rates at very old ages is proposed and validated on both modern and historical data. The author also considers the implications of this model for the distribution of the highest attained age and the question of a possible limit to human lifespan.

We also note the important article by J.Gampe \cite{Gampe}, X. Rootzen and D. Golud \cite{rot_zg}. These authors, relying on real data from the database on the remaining lifespan of supercentenarians (their lifespan after 110 years) and on well-known statistical criteria, came to an important conclusion: the remaining lifespan of a human after 110 years has an exponential distribution  \cite{idl}.

This article also provides some other interesting information about the remaining life expectancy of supercentenarians, as well as a fairly comprehensive bibliography of recent studies on this topic. For example, the article calculates that 47\% of supercentenarians will live another year after age 110. Based on the International Database on Longevity, although there are significantly more women supercentenarians, it is shown that there are no differences between the distributions of remaining life expectancy for women and men supercentenarians. There are also no significant differences in the distributions between the regions of Northern and Southern Europe, between North America and Japan, and other countries. The data were used to calculate the exponential distribution parameter of remaining life expectancy equal to 1.34 and the confidence interval (1.22, 1.46) with a reliability of 0.95.

It is known from studies by \cite{eco} and \cite{grin_i} that, after about 100 years, the human mortality rate practically ceases to increase with age. At first glance, this seems to call into question the use of the Gompertz-Makeham distribution. However, the following calculations for women born in England \& Wales in 1900 over the (80, 120) year interval (sample size $n=154901$) showed that the data are consistent with the Gompertz-Makeham distribution at the traditional significance level of $p=0.05$.

It turns out that the same can be said about the residual life time of men and women in Norway and Sweden in this interval for all the distributions that we consider: GM, GW, LKM, LBM, LBW (see tables \ref{tab:norway1900_80_120},  \ref{tab:sweden1900_80_120}, \ref{tab:england_wales1900_80_120} from section \ref{section6}).

It should be noted that these statements are somewhat surprising given the results of \cite{tha1998}. True, \cite{tha1998} studied pooled data for 13 industrialized countries and a number of cohorts (it is unlikely that such a pooling would yield a homogeneous sample). In our opinion, the calculations presented in \cite{am24} suggest that for aggregated data across several countries and cohorts, simple distributions can only be used after 100 years.

This article attempts to answer the above questions.

Our article is structured as follows. Section \ref{section2} introduces mathematical models that we believe are directly relevant to the task of analyzing residual human life expectancy data. These models will be used in our analysis of real data in Sections \ref{section4} and \ref{section5} (considering the (30, 80) year interval) and Section \ref{section6} (the (80, 120) year interval). In \ref{section3}, we explain the somewhat unexpected appearance of a logistic distribution in the analysis of residual life expectancy data. At the end of the article, we present tables of the grouped data used in the article (Appendix 1, Appendix 2) and provide some calculation examples.

\section{Mathematical models for analyzing mortality data}\label{section2}

As above, $\mathfrak{T}_{t_0}$ is the remaining lifespan of a person after some moment $t_0$, and $ F(x)=
\mathbf{P}(\mathfrak{T}_{t_0}< x) $ is the distribution function of $\mathfrak{T}_{t_0}$ (usually it depends on the value of $t_{0}$).

{\it 1. Gompertz-Makeham (GM) distribution}.

B. Gompertz \cite{gom}, analyzing real data, proposed using an exponentially increasing failure rate function (mortality rate):
\begin{equation} \label{f2.1}
r(x)=\frac{F'(x)}{1-F(x)}= \beta \alpha \exp(\alpha x) , \quad x >0,
\beta >0 , \alpha >0.
\end{equation}

Later, W. Makeham (1860) supplemented formula (\ref{f2.1}) with another parameter $\lambda$:

\begin{equation} \label{f2.2}
 r(x)=\frac{F'(x)}{1-F(x)}= \lambda + \beta \alpha \exp(\alpha x) , \quad x >0,
\beta  >0 , \alpha >0.
 \end{equation}
From formula (\ref{f2.2}) it follows that the distribution function of the value $\mathfrak{T}_{t_0}$ is given by the formula
\begin{equation} \label{f2.3}
  F(x)= F_{GM} (x)= \Bigg\{
 \begin{array}{rl}
   1 - \exp(-\lambda x -{\beta}(\exp(\alpha x) -1)), \quad  \mbox{for} \quad  x\geq 0  \\
  0  , \quad  \mbox{otherwise}.
 \end{array} \Bigg\}
 \end{equation}
We are missing the dependence of the parameters $\lambda$, $\beta$, $\alpha$ on $t_{0} $.

Equalities (\ref{f2.2}), (\ref{f2.3}) mean that mortality is influenced by two components: one that is independent of age (the background component) and the second, which, at $\alpha > 0$, increases exponentially with age (characterizing the wear and tear and aging of the human body).

{\it 2. Weibull distribution (W)}.

Next, we'll consider the Weibull distribution, one of the most popular distributions in reliability theory. Mathematical models of reliability theory have long been used by biologists and gerontologists to analyze mortality data for humans and other living organisms (see \cite{gav-91}, Ch. 6, $\S$ 5, \cite{voy}).

The mortality rate function and the Weibull distribution function are given respectively by the formulas:
\begin{equation} \label{f2.4}
r(x)= \lambda  \gamma x^{\gamma -1},\quad F( x)= F_{W}( x)=1- \exp(-
\lambda x^{\gamma}),  \quad \lambda \geq 0, \quad \gamma \geq
0,\quad x\geq 0.
 \end{equation}
The Weibull distribution is also related to the distribution of extreme values for minima (2nd type according to the classification of B.V. Gnedenko \cite{gne43}, \cite{llr}).

In addition, the importance of this distribution is associated with the following statement from the work \cite{voy}: for 100 major diseases, in 45 cases mortality is better described by the Weibull distribution, and in 55 cases by the Gompertz distribution.

However, some previous calculations using the $\chi^2$ criterion show that the Gompertz-Makeham distribution approximates actual data on total mortality over the (30, 80)-year interval significantly better than the Weibull distribution. Therefore, in this article, the Weibull distribution is used only to describe the exogenous component of mortality (characterizing external influences on a person); more precisely, it is included in the Gompertz-Weibull or Logistic Beard-Weibull distributions.

{\it 3. Gompertz-Weibull distribution (GW)}.

The article \cite{voy} proposes a general function of the intensity of human mortality, presented in the form:
\begin{equation} \label{f2.5}
r(x)=r_1 (x) + r_2 (x),
\end{equation}
where $r_1 (x)$ is the exogenous component of mortality,
and $r_2 (x)$ is the engenogenic component of mortality (it depends on internal processes associated with human aging).

Further, based on well-known models of the mathematical reliability theory, we believe that
$r_1 (x) = \lambda  \gamma x^{\gamma -1}$ and  $r_2 (x) = \beta \alpha \exp(\alpha x)$.
Thus, $\mathfrak{T}_{t_0} =\min (\mathfrak{T}^{'}_{t_0}, \mathfrak{T}^{''}_{t_0})$, where $\mathfrak{T}^{'}_{t_0}$ has a Weibull distribution, and $\mathfrak{T}^{''}_{t_0}$ has a Gompertz distribution, and these variables are independent.

In fact, the formula (\ref{f2.5}) is close in structure to the Gompertz-Makeham formula (\ref{f2.2}), but differs from the latter in that $r_1 (x)$ is not a constant, but depends on age. From here it is not difficult to find the Gompertz-Weibull distribution function:
\begin{equation} \label{f2.6}
  F(x)= F_{GW} (x)= \Bigg\{
 \begin{array}{rl}
   1 - \exp(-\lambda x^{\gamma } -{\beta}(\exp(\alpha x) -1)), \quad  \mbox{for} \quad  x\geq 0  \\
  0  , \quad  \mbox{otherwise} .
 \end{array} \Bigg\}
 \end{equation}

{\it 4. Logistic Beard-Makeham Distribution (LBM)}.

R. E. Beard in his works \cite{bea} and \cite{Beard} considered the following mathematical model for analyzing data about the remaining time of human life. Here it is proposed to set the mortality rate by the formula
\[
 r( x)=\lambda +\frac{a c \exp(bx)}{1+a \exp(bx)},  \quad \lambda \geq 0,
 \quad a \geq 0,\quad b > 0, \quad c > 0,\quad x\geq  0.
 \]
From here we find the distribution function
\begin{equation} \label{f2.7}
  F(x)= F_{LBM} (x)= \Bigg\{
 \begin{array}{rl}
  1- \exp(-
\lambda x)\left(\frac{1+a }{1+a \exp(bx)}\right)^{c/b} , \quad  \mbox{for} \quad  x\geq 0  \\
  0  , \quad  \mbox{otherwise} .
 \end{array} \Bigg\}
 \end{equation}

If we set $c=b$, $\lambda =0$ in formula (\ref{f2.7}), then the function $F_{LBM}(x)$ coincides with the distribution function of $\xi^+$ such that
\[
\mathbf{P}(\xi^+ \geq x) =\mathbf{P}(\xi \geq x|\xi \geq 0),
\]
where the r.v. $\xi$ has the usual logistic distribution
\[
\mathbf{P}(\xi < x)=\frac{1}{1+a
\exp(-bx)}.
\]

Thus, formula (\ref{f2.7}) specifies a certain generalization of the logistic distribution provided that the corresponding r.v. is non-negative.
It should be noted that the article \cite{bea} provides some considerations on how the model (\ref{f2.7}) can arise in a simple heterogeneous population model.

Note that previously a model close to (\ref{f2.7}) was considered by Perks \cite{per}.

By the Logistic Beard-Makeham distribution we will mean the distribution that is given by the formula (\ref{f2.7}).

{\it 5. Logistic Beard-Weibull distribution (LBW)}.
Naturally, by analogy with formulas (\ref{f2.5}) - (\ref{f2.7}), introduce the following distribution with the following mortality intensity:
\[
 r( x)=\lambda  \gamma x^{\gamma -1} +\frac{a c \exp(bx)}{1+a \exp(bx)},  \quad \lambda \geq 0, \quad \gamma > 0, \quad a \geq 0,\quad b > 0,
 \quad c > 0,\quad x\geq
 0,
 \]
with distribution function
\begin{equation} \label{f2.71}
  F(x)= F_{LBW} (x)= \Bigg\{
 \begin{array}{rl}
  1- \exp\left(-\lambda x^{\gamma }\right)
\left(\frac{1+a }{1+a \exp(bx)}\right)^{c/b} , \quad  \mbox{for} \quad  x\geq 0  \\
  0, \quad  \mbox{otherwise} .
 \end{array} \Bigg\}
 \end{equation}
Formula (\ref{f2.71}) defines the Logistic Beard-Weibull distribution.

{\it 6. Logistic Kannisto-Makeham distribution (LKM)}.
In this article we will use another version of the logistic distribution (in the work of \cite{tha1998} it is called the Kannisto distribution).

The $LKM$ distribution has the following mortality intensity (see \cite{am24} , \cite{tha1998}):
\[
 r( x)=\lambda +\frac{a \exp(bx)}{1+a \exp(bx)},  \quad \lambda \geq 0, \quad a \geq 0,\quad b > 0, \quad x\geq 0,
 \]
and the distribution function
\begin{equation} \label{f2.8}
 F_{LKM}( x)=1- \exp\left(-\lambda x\right)\left(\frac{1+a }{1+a \exp(bx) }\right)^{1/b} , \quad x\geq 0.
 \end{equation}
This distribution in the work of \cite{tha1998} is determined as one of the best when analyzing human mortality at the final stage (after 80 years).

It is easy to see that the $LKM$ distribution is a special case of the $LBM$ distribution (it is enough to put $c=1$ in the formula (\ref{f2.7}).

\section{On the extremes of a random number of independent random variables and the logistic distribution.}\label{section3}
If the choice of distributions $GM$, $GW$ and $W$ for the analysis of data on the remaining time of human life is understandable, then questions remain for the choice of the logistic distribution. Especially, if we take into account the critical remarks of Feller (see \cite{fe2}, Ch.2, \S 4) regarding the "transcendental law of logistic development".

In this section, we will attempt to clarify the somewhat "mysterious" appearance of the logistic distribution in the analysis of residual life data. More precisely, here we will present a result from work by B.~V.~Gnedenko  and D.~B.~Gnedenko \cite{BDG82}, which is important for our purposes and which seems little known to the average applied scientist.

Let $ \xi_{nk} = (\xi_{k} -A_n)/B_n$, where $(\xi_{k})$ is a sequence of independent identically distributed random variables (i.i.d.r.v.) with distribution function F(x). The well-known theorem on extreme types \cite{gne43}, \cite{llr} states that the class of possible boundary distributions for $ \max(\xi_{n1}, \xi_{n2},
\ldots, \xi_{nn})$, given an appropriate choice of the normalizing constants $A_n $ and $B_n $, as $ n \rightarrow \infty $, can have only one of the following three types of distributions:
\begin{eqnarray}
&& \mbox{type \, I :}  %\nonumber\\
\quad    \Phi_{1} (x)  =  \exp (-e^{-x})
,\quad \mbox{for} \quad - \infty < x <\infty  . \nonumber\\
&&  \mbox{type \, II :} % \nonumber\\
\quad  \Phi_{2} (x) =  \left \{
 \begin{array}{rl}
 0, & \mbox{for} \quad x \leq 0 \\
 \exp (-x^{-\alpha}), & \mbox{for} \quad  \alpha > 0, \, x > 0 \,
 , \end{array} \right.  \nonumber\\
&& \mbox{type \, III :} % \nonumber\\
\quad  \Phi_{3} (x)  =  \left \{
 \begin{array}{rl}
  \exp (-(-x)^{\alpha}), & \mbox{for} \quad  \alpha > 0, \, x \leq 0  \\
  1, & \mbox{for} \quad x > 0.
 \end{array} \right.  \nonumber\\
\end{eqnarray}

Let the r.v. $\nu_n $ have a geometric distribution with parameter $q_n $, that is, $\mathbf{P}(\nu_n = k) = q_n (1- q_n)^{k-1}$, where $q_n=1/n $. And let the sequence $(\xi_{k})$ be independent of the r.v. $\nu_n $. Then, by the second theorem from the article \cite{BDG82}, the possible boundary distributions for
\[\max(\xi_{k} - A_n : 1 \leq k \leq \nu_n )/B_n \]
with the same choice of normalizing constants $A_n$, $B_n $ are limited to only the following types:
\begin{eqnarray}
&& \mbox{type \, I :} % \nonumber\\
 \quad L(x)  = L_{1} (x)  =  1/(1+ e^{-x})
,\quad \mbox{for} \quad - \infty < x <\infty  . \nonumber\\
 && \mbox{type \, II :} %  \nonumber\\
\quad  L_{2} (x) =  \left \{
 \begin{array}{rl}
 0, & \mbox{for} \quad x \leq 0 \\
 1/(1+ x^{-\alpha}), & \mbox{for} \quad  \alpha > 0, \, x > 0 \,
 , \end{array} \right.  \nonumber\\
&& \mbox{type \, III :} %  \nonumber\\
\quad L_{3} (x)  =  \left \{
 \begin{array}{rl}
  1/(1+ (-x)^{\alpha}), & \mbox{for} \quad  \alpha > 0, \, x \leq 0  \\
  1, & \mbox{for} \quad x > 0.
 \end{array} \right.  \nonumber\\
\end{eqnarray}

The distribution $L(x)$ is the logistic distribution.

It is not difficult to move from the maximum scheme to the minimum scheme, which is what interests us most. To do this, we'll use a simple equality:
\[
\min_{1\leq i \leq n} \xi_i = -\max_{1\leq i \leq n}(- \xi_i ).
\]

Suppose that for $n\rightarrow \infty$
\[
\mathbf{P}(\max(-\xi_{k} - A_n : 1 \leq k \leq n )/B_n < x) \rightarrow \Phi_{1} (x) .
\]
Then, according to the above statement, we have
\[
\mathbf{P}(\max(-\xi_{k} - A_n : 1 \leq k \leq \nu_n )/B_n < x) \rightarrow L(x) .
\]
Therefore,
\[
\mathbf{P}(\min(\xi_{k} + A_n : 1 \leq k \leq \nu_n )/B_n \geq -x) \rightarrow L(x) .
\]

From this we have
\[
\mathbf{P}(\min(\xi_{k} + A_n : 1 \leq k \leq \nu_n )/B_n <x) \rightarrow 1 -L(-x)= L(x).
\]
It is known \cite{BDG82} that convergence to $ \Phi_{1} (x) $ for the maximum scheme will be observed under sufficiently broad conditions. This circumstance and the above results from the article \cite{BDG82} demonstrate the special role of the logistic distribution (as well as its generalizations $LBW$ and $LBM$ distributions) in the analysis of data on the remaining time of human life.

\section{Statistical analysis of data on the remaining time of human life over the interval (30,80) years.}\label{section4}

Let $(t_0, t_r)$ be some interval, $0<t_0 < t_r < \infty,
F(x)$ be the distribution function, $ F(0) =0, $
$ \mathfrak{T}_{t_0}$ be the remaining lifespan of a person who has reached the age $t_0$.

Let us put forward the hypothesis:
\[
H(t_0, t_r) =\{\mathbf{P} (\mathfrak{T}_{t_0}< t) =F(t-t_0), \quad \forall t \in (t_0, t_r)\},
\]
that is, the assumption that the r.v. $ \mathfrak{T}_{t_0}$ on the interval $(t_0, t_r)$ has the distribution $F(t-t_0)$.

Our efforts in this section will primarily focus on testing the hypothesis: $H(t_0, t_r)$ for $t_0 = 30$, $t_r = 80$ in the case of the $GM$, $GW$, $LKM$, $LBM$ and $LBW$ distributions introduced above. Unfortunately, we cannot use statistical tests such as Kolmogorov-Smirnov, Cramer-Von Mises $\omega^2$, and the like (since the critical points for our distributions are unknown when estimating unknown parameters, and we will also have grouped data). The well-known $\chi^2$ criterion, \cite{cra}, will be our primary criterion.

Let us assume that in some region the grouped data on mortality of people $n_1$, $n_2, \ldots$, $n_r , n_{r+1} $ are known on the corresponding intervals $(t_0, t_1)$, $(t_1, t_2), \ldots$, $(t_{r-1}, t_r)$, $(t_{r} , t_{r+1})$, where $N= n_1 + n_2 + \ldots + n_r + n_{r+1} $ is the number of all people in this region who have reached the age of $t_0$. For the specific case of the hypothesis $H(t_0, t_r)$ with $t_0 = 30, \, t_r = 80$, we select the intervals as follows:
$(30, 31), (31, 32), \ldots , (79, 80), (80, 120)$.

Next, for example, we will consider $GM$ distribution. Let's introduce the statistics $\chi^2$:
\[
\chi^2(\lambda, \beta, \alpha) = \sum_{i=1}^{r+1} \frac{(n_i -N\cdot
p_i(\lambda, \beta, \alpha))^2}{N \cdot p_i(\lambda, \beta, \alpha)} \approx \sum_{i=1}^{r+1} \frac{(n_i -N\cdot
p_i(\lambda, \beta, \alpha))^2}{n_i},
\]
where $p_i(\lambda, \beta, \alpha) =F_{GM}(\lambda, \beta, \alpha, t_i -t_0)
- F_{GM}(\lambda, \beta, \alpha, t_{i-1} -t_0)$ and $F_{GM} $
is given by equality (\ref{f2.3}).

Just as in the article \cite{am24} (see also \cite{cra}, Ch. 30, \S 3 ),
to test the hypothesis $ H(t_0, t_r)$, we use the method of minimum statistic $\chi^2$:
\begin{equation} \label{f3.1}
\chi^2_{min}= \chi^2(\lambda^*, \beta^*, \alpha^*) = \min_{\lambda >0, \beta>0, \alpha>0
} \chi^2(\lambda, \beta, \alpha),
\end{equation}
that is, to find the optimal parameters $\lambda^*, \beta^*, \alpha^*$ of the $GM$ distribution, it is necessary to solve the minimization problem (\ref{f3.1}).

In this case, it will be essential to use the functions of the Optimization Toolbox package of the MATLAB system (see article \cite{am24}). Note that the problem (\ref{f3.1}) is far from trivial and in many cases it is important to successfully choose the starting point from which the minimization algorithm starts.

The $\chi^2$ statistics for the case of $GW, LKM, LBM$ and $LBW$ distributions are introduced and analyzed in a similar way.

Hypothesis testing for $ H(t_0, t_r)$ follows the standard procedure. If $\chi^2_{\min} > x_{kr} $, then $ H(t_0, t_r)$ is rejected. Otherwise, we assume that the data do not contradict the hypothesis
$ H(t_0, t_r)$, where $x_{kr} =U_{0.95} $ is the ${0.95}^{th}$ quantile of the $\chi^2$ distribution with $r+1-m-1$ degrees of freedom, $m$ is the number of parameters of the corresponding distribution, and $\chi^2_{\min} $ is the minimum value of the $\chi^2$ statistic obtained from problem (\ref{f3.1}).

In this article we deal with samples of large size $n$. For example, for women England \& Wales $n=340450$. In such a case, the hypothesis $ H(t_0, t_r)$ for $t_0 =30$, $t_r =80$ is often rejected at the significance level $p=0.05$ and the distributions are compared by the value $\chi^2_{\min} $.

It should be noted that during calculations in the MATLAB system, the following designations of unknown parameters were used: \\
$GM$-distribution (formula (\ref{f2.3})), $\lambda = x(3)^2$, $\beta = x(1)^2$, $\alpha = x(2)^2$;\\
$GW$-distribution (formula (\ref{f2.6})), $\lambda = x(3)^2$, $\beta = x(1)^2$, $\alpha = x(2)^2$, $\gamma = x(4)^2$ ;\\
$LKM$-distribution (formula (\ref{f2.8})), $\lambda = x(3)^2$, $a = x(1)^2$, $b = x(2)^2$;\\
$LBM$-distribution (formula (\ref{f2.7})), $\lambda = x(3)^2$, $a = x(1)^2$, $b = x(2)^2$, $= x(4)^2$;\\
$LBW$-distribution (formula (\ref{f2.71})), $\lambda = x(3)^2$, $a = x(1)^2$, $b = x(2)^2$, $c= x(4)^2$, $\gamma =x(5)^2$.

In addition, the following designations are used here and further: \\
$(t_0, t_r +)$ is the age interval on which the final time of human life is analyzed; \\
$N(t_0)$ - the number of people who have reached the age of $t_0$;\\
$\chi^2_{min}$ - the minimum value of the statistic $\chi^2$ for the corresponding distribution;\\
${x}^*$ - optimal values of the parameters at which $\chi^2_{min}$ is reached;\\
$r-m$ - the number of degrees of freedom in the $\chi^2$ criterion, $r+1$ - the number of grouping intervals;\\
$m$ - the number of unknown parameters of the corresponding distribution; \\
$U_{0.95}$ is the 0.95 level quantile of the $\chi^2$ distribution with $r-m$ degrees of freedom.

\vspace{0.5cm}

{\it Example 1. Norway: cohort born in 1900, interval (30, 80+) years.}\\
Grouped data on mortality of the population of Norway after 30 years (cohort born in 1900) are given in Appendix 1.

The following Table \ref{tab:norway1900} contains the results of calculations of the minimum values of $\chi^2$ statistics for mortality data separately for men and women in Norway over the interval (30, 80+) years.

\begin{table}[h]
\centering
\small

\caption{Norway, cohort born in 1900, interval (30, 80+) years}
\label{tab:norway1900}

\resizebox{\textwidth}{!}{
\begin{tabular}{|l|c|c|}
\hline
\textbf{Gender} & \textbf{Women} & \textbf{Men} \\
\hline\hline

$(t_0 , t_{r}^{+})$ 
& $(30 , 80+)$ 
& $(30 , 80+)$ \\
\hline

$N(t_0)$ 
& 23208 
& 22450 \\
\hline\hline

\textbf{GM} 
& $\chi^2_{min} = 57.39$ 
& $\chi^2_{min} = 155.66$ \\
\cline{2-3}
& $x^* =(0.0286, 0.3582, 0.0552)$ 
& $x^* =(0.0576, 0.3362, 0.0580)$ \\
\hline\hline

\textbf{GW} 
& $\chi^2_{min} = 31.54$ 
& $\chi^2_{min} = 61.41$ \\
\cline{2-3}
& $x^* =(0.0348, 0.3485, 0.0674, 0.9249)$ 
& $x^* =(3.3042, 0.0203, 0.0000, 2.3356)$ \\
\hline\hline

\textbf{LKM} 
& $\chi^2_{min} = 52.90$ 
& $\chi^2_{min} = 137.15$ \\
\cline{2-3}
& $x^* =(0.0097, 0.3627, 0.0555)$ 
& $x^* =(0.0179, 0.03433, 0.0590)$ \\
\hline\hline

\textbf{LBM} 
& $\chi^2_{min} = 35.52$ 
& $\chi^2_{min} = 49.40$ \\
\cline{2-3}
& $x^* =(0.0177, 0.3961, 0.0574, 0.3589)$ 
& $x^* =(0.0239, 0.4093, 0.0655, 0.3445)$ \\
\hline\hline

\textbf{LBW} 
& $\chi^2_{min} = 22.90$ 
& $\chi^2_{min} = 49.41$ \\
\cline{2-3}
& $x^* =(0.0200, 0.3727, 0.0664, 0.4510, 0.9438)$ 
& $x^* =(0.0243, 0.4078, 0.0662, 0.3463, 0.9957)$ \\
\hline

$r-m-1$ 
& 51-5-1 = 45  
& 51-5-1 = 45  \\
\hline

$U_{0.95}$ 
& 61.65 (64 for $m=3$) 
& 61.65 (64 for $m=3$) \\
\hline

\end{tabular}
}

\noindent\textbf{Source:} Author's calculations based on data from the Human Mortality Database. 
Available at: \url{https://www.mortality.org/Country/Country?cntr=NOR}. 
Accessed: 2026-03-15.

\end{table}

Comments on the Table \ref{tab:norway1900}:

(i) According to Table \ref{tab:norway1900}, for Norwegian women, the $\chi^2$ test accepted the $H(30, 80+)$ hypothesis at a significance level of $p=0.05$ for all distributions: $GM$, $GW$, $LKM$, $LBM$, $LBW$.

(ii) A similar situation holds for Norwegian men in the case of the $GW$, $LBM$, and $LBW$ distributions.

(iii) At the same time, the $\chi^2$ test data for Norwegian men contradict the $GM$ and $LKM$ distributions even at a significance level of $p=5\cdot 10^{-4}$.

(i$\vee)$ The $LBM$ and $LBW$ distributions provided the best approximation to the actual data for the $\chi^2_{min}$ value, while the $GM$ distribution had the worst indicator.

{\it Remark 1.} Unfortunately, in some cases, the influence of certain external factors on the data significantly alters them, rendering the analysis inadequate. For example, if we analyze mortality data for Norwegian women born in 1881 over the age interval (30, 80) similarly to Table \ref{tab:norway1900}, we obtain a completely different picture. The hypothesis $H(30, 80)$ will be rejected by the $\chi^2$ criterion at the significance level $p=0.05$ for all distributions $GM$, $GW$, $LKM$, $LBM$, $LBW$.

\begin{figure}
    \centering
    \includegraphics[width=1\linewidth]{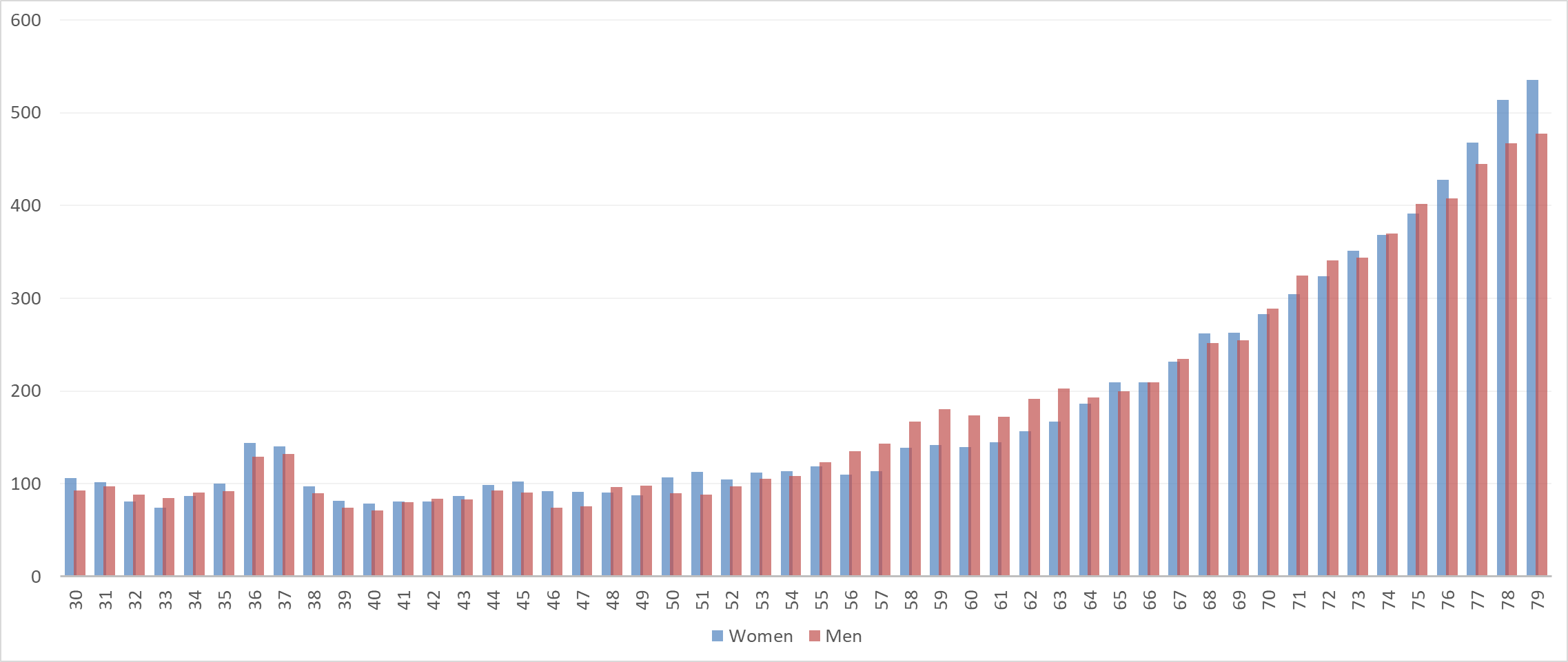}
       \caption{Empirical histogram of mortality in Norwegian women and men born in 1881 in the interval (30, 80 years).}
    \label{fig1}
\end{figure}

The following explanation seems plausible. Graphic analysis of the empirical histogram shows (Figure \ref{fig1}) that over the interval (34, 38) years there is a significant increase in mortality (for both women and men). But in fact, this interval corresponds to the years of the 1st World War (1914-1918), which gives such a result.

Another confirmation of this argument is the following. Calculations show that for the same data on the interval of (40, 90) years, the hypothesis $H(40, 90)$ is accepted for all distributions given above.

A similar remark applies to Norwegian men born in 1881

{\it Example 2. Sweden, cohort born in 1900, interval (30, 80+) years.}\\
Grouped data on mortality of the population of Sweden after 30 years (cohort born in 1900) are given in Appendix 1.

The following Table \ref{tab:sweden1900} contains the results of calculations of the minimum values of $\chi^2$ statistics for mortality data separately for men and women in Sweden over the interval (30, 80+) years.

\begin{table}[h]
\centering
\small

\caption{Sweden: cohort born in 1900, interval (30, 80+) years}
\label{tab:sweden1900}

\resizebox{\textwidth}{!}{
\begin{tabular}{|l|c|c|}
\hline
\textbf{Gender} & \textbf{Women} & \textbf{Men} \\
\hline\hline

$(t_0 , t_{r}^{+})$ 
& $(30 , 80+)$ 
& $(30 , 80+)$ \\
\hline

$N(t_0)$ 
& 51352 
& 50000 \\
\hline\hline

\textbf{GM} 
& $\chi^2_{min} = 75.48$ 
& $\chi^2_{min} = 164.65$ \\
\cline{2-3}
& $x^* =(0.0412, 0.3386, 0.0529)$ 
& $x^* =(0.0621, 0.3318, 0.0502)$ \\
\hline\hline

\textbf{GW} 
& $\chi^2_{min} = 58.09$ 
& $\chi^2_{min} = 67.81$ \\
\cline{2-3}
& $x^* =(0.0469, 0.3319, 0.0598, 0.9498)$ 
& $x^* =(3.3090, 0.0202, 0.0000, 2.3356)$ \\
\hline\hline

\textbf{LKM} 
& $\chi^2_{min} = 73.20$ 
& $\chi^2_{min} = 140.58$ \\
\cline{2-3}
& $x^* =(0.0133, 0.3427, 0.0533)$ 
& $x^* =(0.0192, 0.3386, 0.0514)$ \\
\hline\hline

\textbf{LBM} 
& $\chi^2_{min} = 71.28$ 
& $\chi^2_{min} = 87.04$ \\
\cline{2-3}
& $x^* =(0.0209, 0.3512, 0.0541, 0.5799)$ 
& $x^* =(0.0310, 0.3699, 0.0557, 0.4403)$ \\
\hline\hline

\textbf{LBW} 
& $\chi^2_{min} = 58.31$ 
& $\chi^2_{min} = 61.57$ \\
\cline{2-3}
& $x^* =(0.0146, 0.3363, 0.0597, 1.0141, 0.9540)$ 
& $x^* =(0.0336, 0.3547, 0.0641, 0.4997, 0.9338)$ \\
\hline

$r-m-1$ 
& 51-5-1 = 45  
& 51-5-1 = 45  \\
\hline

$U_{0.95}$ 
& 61.65 (64 for $m=3$) 
& 61.65 (64 for $m=3$) \\
\hline

\end{tabular}
}

\noindent\textbf{Source:} Author's calculations based on data from the Human Mortality Database. 
Available at: \url{https://www.mortality.org/Country/Country?cntr=SWE}. 
Accessed: 2026-03-15.

\end{table}

Comments on the Table \ref{tab:sweden1900}:

(i) From Table \ref{tab:sweden1900}, we conclude that for Swedish women, the $\chi^2$ criterion accepted the $H(30, 80)$ hypothesis at a significance level of $p=0.05$ for the $GW$, $LBW$ distributions.

(ii) For Swedish men, the optimal $\chi^2_{min}$ values were higher than for women, and the $H(30, 80)$ hypothesis was accepted only for the $LBW$ distribution.

(iii) Moreover, the mortality data for Swedish men over the (30, 80+)-year interval contradict the $GM$ and $LKM$ distributions even at a significance level of $p=5\cdot 10^{-4}$.

(i$\vee)$ The $GW$ and $LBW$ distributions provided the best approximation to the actual data for Sweden over the (30, 80+)-year interval, while the $GM$ distribution again had the worst value.

It should be noted that the sample sizes for Sweden are more than twice those for Norway.

\vspace{2mm}
{\it Example 3. England \textnormal{\&} Wales: cohort born in 1900, interval (30, 80+) years.}\\
Grouped data on mortality of the population of England \& Wales after 30 years (cohort born in 1900) are given in Appendix 1.

The following Table \ref{tab:engwales1900} contains the results of calculations of the minimum values of $\chi^2$ statistics for mortality data separately for men and women in England \& Wales over the interval (30, 80+) years.

\begin{table}[h]
\centering
\small

\caption{England \& Wales, cohort born in 1900, interval (30, 80+) years}
\label{tab:engwales1900}

\resizebox{\textwidth}{!}{
\begin{tabular}{|l|c|c|}
\hline
\textbf{Gender} & \textbf{Women} & \textbf{Men} \\
\hline\hline

$(t_0 , t_{r}^{+})$ 
& $(30 , 80+)$ 
& $(30 , 80+)$ \\
\hline

$N(t_0)$ 
& 340450 
& 323759 \\
\hline\hline

\textbf{GM} 
& $\chi^2_{min} = 161.00$ 
& $\chi^2_{min} = 1127$ \\
\cline{2-3}
& $x^* =(0.0500, 0.3297, 0.0521)$ 
& $x^* =(0.1146, 0.3055, 0.0388)$ \\
\hline\hline

\textbf{GW} 
& $\chi^2_{min} = 135.75$ 
& $\chi^2_{min} = 138.48$ \\
\cline{2-3}
& $x^* =(0.0565, 0.3268, 0.0550, 0.9755)$ 
& $x^* =(0.4398, 0.1338, 0.0001, 2.1769)$ \\
\hline\hline

\textbf{LKM} 
& $\chi^2_{min} = 128.57$ 
& $\chi^2_{min} = 855$ \\
\cline{2-3}
& $x^* =(0.0168, 0.3345, 0.0527)$ 
& $x^* =(0.0325, 0.3139, 0.0421)$ \\
\hline\hline

\textbf{LBM} 
& $\chi^2_{min} = 88.11$ 
& $\chi^2_{min} = 212.30$ \\
\cline{2-3}
& $x^* =(0.0278, 0.3482, 0.0543, 0.5216)$ 
& $x^* =(0.0507, 0.3533, 0.0528, 0.4406)$ \\
\hline\hline

\textbf{LBW} 
& $\chi^2_{min} = 87.41$ 
& $\chi^2_{min} = 205.67$ \\
\cline{2-3}
& $x^* =(0.0276, 0.3466, 0.0548, 0.5356, 0.9955)$ 
& $x^* =(0.0519, 0.3501, 0.0542, 0.4486, 0.9838)$ \\
\hline

$r-m-1$ 
& 51-5-1 = 45  
& 51-5-1 = 45 \\
\hline

$U_{0.95}$ 
& 61.65 (64 for $m=3$) 
& 61.65 (64 for $m=3$) \\
\hline

\end{tabular}
}
\noindent\textbf{Source:} Author's calculations based on data from the Human Mortality Database. 
Available at: \url{https://www.mortality.org/Country/Country?cntr=GBRCENW}. 
Accessed: 2026-03-15.

\end{table}

Comments on the Table \ref{tab:engwales1900}:

(i) At the significance level $p=0.05$, the mortality data for both women and men in England \& Wales are inconsistent with the distributions of $GM$, $GW$, $LKM$, $LBM$, $LBW$. Moreover, the minimum values of $\chi^2_{min}$ for all distributions exceed the critical point even at the significance level $p=5\cdot 10^{-4}$. The best approximation to the replicate data over the interval (30, 80+) years for the $\chi^2$ criterion again yielded the $LBW$ distribution.

(ii) Note that the sample size for women mortality in England \& Wales is $n=340450$, more than 14 times larger than the sample size for Norway. Furthermore, in real-world situations, a number of unforeseen deterministic factors (e.g., wars, epidemics, immigration, emigration, etc.) significantly influence the data we analyze. We are unaware of any non-trivial examples of real-world data of this size that would adequately describe the data with a simple distribution at a significance level of $p=0.05$.

(iii) Since the $\chi^2$ criterion rejected the distributions proposed in section \ref{section2}, the question arises: how to logically describe the data in this case?

Next, we propose one approach to analyzing large-volume real data, which can be useful in the case of mortality data of the England \& Wales type.

\section{Using the "thinning" procedure in analyzing mortality data for England \& Wales.
Interval (30, 80) years.}\label{section5}

It is logical to assume that in the ''ideal'' case, the England \& Wales population mortality data over the interval (30, 80) years are adequately described by the $LBW$ distribution. But, as noted above, in reality, certain reasons (for example, war or immigration) significantly changed these data.

Thus, for the data in Table \ref{tab:engwales1900}, even for the best $LBW$ distribution for both women and men, the optimal values of the terms $d_i^2 =(n_i -N p_i)^2 /(N p_i)$ of the $\chi^2$ statistic over the interval $(40, 41)$ years exceeded the critical point $x^2_{up}$, $x_{up}\approx 4.7$, for which $ P-value = \mathbf{P}(\max_{1\leq i \leq 51} \xi_i > x_{up})=5 \cdot 10^{-5}$, $\xi_i$ are independent standard distributed r.v.

At the same time, for the populations of Norway and Sweden in the interval $(40, 41)$ years, no such excess of the critical point $x^2_{up}$ was observed. In our opinion, this means that the results of the test of the $H(30, 80)$ hypothesis, even for the civilian population of England \& Wales, were significantly influenced by World War II.

It is important to note that in the interval $ (62, 63)$ years, the optimal values of the terms $ d_i^2$ exceeded the critical point $ x^2_{up}$, $ x_{up}\approx 4.27$, $ P-value = 5\cdot 10^{-4}$ (for men) and $ x_{up}\approx 3.82$, $ P-value = 5\cdot 10^{-3}$ (for women).

It is known that at the end of 1961 - at the beginning of 1962, England experienced a massive wave of immigration from Commonwealth countries in connection with the Commonwealth Immigrants Bill law \cite{hansard1961}.

Taking into account the above exceedances of critical points, it is clear that the estimates of the distribution parameters $LBW$ and others obtained in Table \ref{tab:engwales1900} will be unsatisfactory.

The question arises: can these estimates be improved using real data? How can this be done?

We will assume that the relevant samples are sufficiently large (for example, as in the case of the populations of England \& Wales or Sweden). Furthermore, it is assumed that, at most grouping intervals, real observations are weakly dependent on unforeseen deterministic factors.

Under these conditions, the following algorithm for estimating unknown distribution parameters is proposed.

1) Let us assume a grouped sample: $(t_0, t_1), (t_1, t_2), \ldots , (t_{r-1}, t_r), (t_{r}, t_{r+1})$ are the grouping intervals, $n_1, n_2, \ldots , n_r, n_{r+1}$ are the number of deaths in the corresponding intervals.

2) The "thinning" operation was performed on this sample, where each element of the sample with probability ($1-q$) is removed from further consideration (e.g., $q = 0.1$ or $q = 0.5$). In this case, new grouped data are obtained on the same grouping intervals - $n^{'}_1, n^{'}_2, \ldots , n^{'}_r , n^{'}_{r+1} $.

3) It is assumed that the removal operation is carried out for each point independently using a random number sensor.

4) In a real situation, after using this operation, the sample becomes significantly smaller, and at the same time, the influence of unforeseen deterministic factors on the assessment results also decreases significantly.

5) Using the new sparse data, just like above, the minimum $\chi^2$ method is used to estimate the distribution parameters and analyze the results.

6) To select the optimal value of the parameter $q$, it is proposed to successively decrease it. For example, we choose $q= 1/2,\, 1/4,\, 1/8, \ldots $ until $\chi^2_{min}$ decreases significantly. Furthermore, the new values of the grouped data ($n^{'}_i, i=1, 2,\ldots, {r+1} $) should not be too small ($n^{'}_i > 15 , \, i=1, 2,\ldots, {r+1}$).

In fact, if the last condition is satisfied, then the adjustment is made on the value of $q$ for which $\chi^2_{min}$ will be minimal.

{\it Remark 2.} Let's provide some explanations regarding this algorithm.

a) If we have an "ideal" sample, that is, a sequence of independent observations of some random variable that has an $LBW$ distribution, then as a result of the "thinning" operation we obtain a sample of size $N^{'} \approx q \cdot N$ also with an $LBW$ distribution (though $N^{'}$ is random and depends only on $N$ and $q$).

b) If, as a result of the action of some unforeseen factor, the mortality rate increased, for example, in the interval $(40, 41)$ year, then the number of points in this interval increased. Therefore, it will decrease accordingly at some subsequent grouping intervals. Therefore, with thinning, the number of removed points on the interval of $ (40, 41)$ years will increase on average, and at some subsequent intervals it will decrease.

c) As a consequence of items a) and b) we have: as a result of "thinning", the influence of unforeseen factors on estimates of the distribution parameters $LBW$ will decrease (just compare tables \ref{tab:engwales1900} and \ref{tab:engwales1900_thinned}).

An example of using this algorithm for population mortality data in England \& Wales, born in 1900, shown in Table \ref{tab:engwales1900_thinned}.

\begin{table}[h]
\centering
\small

\caption{England \& Wales, cohort born in 1900, interval (30, 80+) years, thinning applied ($q=0.1$ for women, $q=0.02$ for men)}
\label{tab:engwales1900_thinned}

\resizebox{\textwidth}{!}{
\begin{tabular}{|l|c|c|}
\hline
\textbf{Gender} & \textbf{Women ($q = 0.1$)} & \textbf{Men ($q = 0.02$)} \\
\hline\hline

$(t_0 , t_{r}^{+})$ 
& $(30 , 80+)$ 
& $(30 , 80+)$ \\
\hline

$N(t_0)$ 
& 34116 
& 6466 \\
\hline\hline

\textbf{GM} 
& $\chi^2_{min} = 56.14$ 
& $\chi^2_{min} = 63.74$ \\
\cline{2-3}
& $x^* =(0.0547, 0.3283, 0.0511)$ 
& $x^* =(0.1146, 0.3051, 0.0413)$ \\
\hline\hline

\textbf{GW} 
& $\chi^2_{min} = 54.72$ 
& $\chi^2_{min} = 59.94$ \\
\cline{2-3}
& $x^* =(0.0571, 0.3261, 0.0532, 0.9816)$ 
& $x^* =(0.1237, 0.3008, 0.0477, 0.8486)$ \\
\hline\hline

\textbf{LKM} 
& $\chi^2_{min} = 52.56$ 
& $\chi^2_{min} = 59.62$ \\
\cline{2-3}
& $x^* =(0.0171, 0.3331, 0.0517)$ 
& $x^* =(0.0328, 0.3128, 0.0399)$ \\
\hline\hline

\textbf{LBM} 
& $\chi^2_{min} = 47.37$ 
& $\chi^2_{min} = 50.08$ \\
\cline{2-3}
& $x^* =(0.0292, 0.3484, 0.0535, 0.4979)$ 
& $x^* =(0.0520, 0.3458, 0.0496, 0.4636)$ \\
\hline\hline

\textbf{LBW} 
& $\chi^2_{min} = 47.27$ 
& $\chi^2_{min} = 46.18$ \\
\cline{2-3}
& $x^* =(0.0292, 0.3503, 0.0530, 0.4846, 1.0052)$ 
& $x^* =(1.5385, 0.4967, 0.0001, 0.0644, 2.1399)$ \\
\hline

$r-m-1$ 
& 51-5-1 = 45  
& 51-5-1 = 45 \\
\hline

$U_{0.95}$ 
& 61.65 (64 for $m=3$) 
& 61.65 (64 for $m=3$) \\
\hline

\end{tabular}
}

\noindent\textbf{Source:} Author's calculations based on data from the Human Mortality Database (England \& Wales, cohort 1900), with thinning applied.

\end{table}

Comments on the Table \ref{tab:engwales1900_thinned}:

(i) For thinning data on mortality of women in England \& Wales with $q=0.1$, the $\chi^2$ test at the significance level $p=0.05$ accepted the hypothesis $H(30, 80)$ for all distributions $GM$, $GW$, $LKM$, $LBM$, $LBW$. The best approximations over the interval (30, 80) years gave the distributions $LBM$, $LBW$.

(ii) In the case of mortality data for men in England \& Wales with a thinning parameter of $q=0.02$, the $\chi^2$ criterion1 accepted the hypothesis $H(30, 80)$ for the distributions of $GW$, $LKM$, $LBM$, $LBW$ at the significance level of $p=0.05$. In the case of the $ GM $ distribution, the value $\chi^2_{min}$ fell on the boundary of the critical region. The best distributions again turned out to be $LBM$, $LBW$.

(iii) When testing the hypothesis $H(30, 80)$, the $\chi^2_{min}$ value for men significantly exceeded the corresponding values for women. A similar situation was observed for sparse mortality data in England \& Wales. For example, in the previous calculations with $q=0.1$, the $\chi^2$ test at a significance level of $p=0.05$ rejected the hypothesis $H(30, 80)$ for the $GM$, $GW$, $LKM$, and $LBM$ distributions for men and accepted it for women. Therefore, in Table 4, a transition was made from $q=0.1$ to $q=0.02$ for men, which significantly reduced the $\chi^2_{min}$ value for all distributions.

{\it Remark 3.} If our assumptions are not met, and for example, the influence of unforeseen deterministic factors at most intervals significantly changes the mortality rate, then the given algorithm will most likely not give positive results.

In this case, we can only repeat the words of J. Neumann: ''Statistics are powerless against corrupted input data''.

\section{Statistical analysis of data for the remaining time of human life. Interval (80, 120) years.}\label{section6}

Let $t_0 =80$ years, $ \mathfrak{T}_{t_0}= \mathfrak{T}_{80}$ be the remaining life time of a person who has reached the age of $80$ years, $ F(x)$ be a function of one of the distributions $GM$, $GW$, $LKM$, $LBM$, $LBW$.

Consider the hypothesis:
\[
H(80, 120) =\{\mathbf{P} (\mathfrak{T}_{80}< t) =F(t-80), \quad \forall t \in [80, 120)\}.
\]
In fact, this section will test the hypothesis that the remaining human lifespan $\mathfrak{T}_{80}$ has the distribution $F(t-80), t \geq 80$. Note that the International Database on Longevity only records one person who lived more than 120 years \cite{idl}.

The method for testing the hypothesis $H(80, 120)$ is similar to the methods in the previous sections 4 and 5. The only slight difference is in the selection of grouping intervals. First, we select them as follows: $(80, 81), (81, 82), \ldots , (119, 120), (120, \infty)$. Next, the final intervals are combined so that each interval contains at least 15 or 20 points. This yields the desired grouping intervals: $(80, 81), \ldots , (t_{r-1}, t_r), (t_r, \infty)$, $t_r < 120$.

In the same way, as above, the minimum values of statistics $\chi^2$ are calculated for the remaining life expectancy separately for men and women after 80 years.

\vspace{0.5cm} 
{\it Example 4. Norway: cohort born in 1900, interval (80, 120) years.}

The following Table \ref{tab:norway1900_80_120} contains the results of calculations of the minimum values of $\chi^2$ statistics for mortality data separately for women and men in Norway over the interval (80, 120) years.

\begin{table}[h]
\centering
\small

\caption{Norway, cohort born in 1900, interval (80, 120) years}
\label{tab:norway1900_80_120}

\resizebox{\textwidth}{!}{
\begin{tabular}{|l|c|c|}
\hline
\textbf{Gender} & \textbf{Women} & \textbf{Men} \\
\hline\hline

$(t_0 , t_{r})$ 
& $(80 , 120)$ 
& $(80 , 120)$ \\
\hline

$N(t_0)$ 
& 12084 
& 7434 \\
\hline\hline

\textbf{GM} 
& $\chi^2_{min} = 28.55$ 
& $\chi^2_{min} = 23.56$ \\
\cline{2-3}
& $x^* =(0.7657, 0.3255, 0.0000)$ 
& $x^* =(1.0322, 0.3018, 0.0012)$ \\
\hline\hline

\textbf{GW} 
& $\chi^2_{min} = 25.97$ 
& $\chi^2_{min} = 22.58$ \\
\cline{2-3}
& $x^* =(0.7579, 0.3218, 0.0269, 1.4036)$ 
& $x^* =(1.0184, 0.3041, 0.0665, 0.0049)$ \\
\hline\hline

\textbf{LKM} 
& $\chi^2_{min} = 19.97$ 
& $\chi^2_{min} = 22.85$ \\
\cline{2-3}
& $x^* =(0.2221, 0.3758, 0.1190)$ 
& $x^* =(0.3153, 0.3405, 0.0626)$ \\
\hline\hline

\textbf{LBM} 
& $\chi^2_{min} = 19.64$ 
& $\chi^2_{min} = 20.52$ \\
\cline{2-3}
& $x^* =(0.2248, 0.3947, 0.1458, 0.8987)$ 
& $x^* =(0.2825, 0.4196, 0.2329, 0.7768)$ \\
\hline\hline

\textbf{LBW} 
& $\chi^2_{min} = 19.64$ 
& $\chi^2_{min} = 20.71$ \\
\cline{2-3}
& $x^* =(1.6353, 2.8535, 0.0285, 0.2605, 1.6505)$ 
& $x^* =(5.2484, 5.6668, 0.0409, 0.3198, 1.5909)$ \\
\hline

$r-m-1$ 
& 25-5-1 = 19 
& 22-5-1 = 16 \\
\hline

$U_{0.95}$ 
& 30.14 (32.67 for $m=3$) 
& 26.3 (28.87 for $m=3$) \\
\hline

\end{tabular}
}

\noindent\textbf{Source:} Author's calculations based on data from the Human Mortality Database. 
Available at: \url{https://www.mortality.org/Country/Country?cntr=NOR}. 
Accessed: 2026-03-15.
\end{table}

Comments on the Table \ref{tab:norway1900_80_120}:

Here $t_0=80, t_{r+1}=\infty$, $(t_r, \infty)$ is the last grouping interval at which the remaining lifetime is analyzed. Other notations are similar to those given above in Section 4.

(i) According to Table \ref{tab:norway1900_80_120}, the $\chi^2$ criterion accepted the $H(80, 120)$ hypothesis at the significance level of $p=0.05$ for all distributions: $GM$, $GW$, $LKM$, $LBM$, $LBW$ for both women and men in Norway.

(ii) Given the different number of parameters in the given distributions, it seems that there is no significant difference when using them to analyze the mortality rate of the Norwegian population over the interval (80, 120) years.

\vspace{0.5cm}

{\it Example 5. Sweden: cohort born in 1900, interval (80, 120) years.}

The following Table \ref{tab:sweden1900_80_120} contains the results of calculations of the minimum values of $\chi^2$ statistics for mortality data separately for women and men in Sweden over the interval (80, 120) years.

\begin{table}[h]
\centering
\small

\caption{Sweden, cohort born in 1900, interval (80, 120) years}
\label{tab:sweden1900_80_120}

\resizebox{\textwidth}{!}{
\begin{tabular}{|l|c|c|}
\hline
\textbf{Gender} & \textbf{Women} & \textbf{Men} \\
\hline\hline

$(t_0 , t_{r})$ 
& $(80 , 120)$ 
& $(80 , 120)$ \\
\hline

$N(t_0)$ 
& 26432 
& 17220 \\
\hline\hline

\textbf{GM} 
& $\chi^2_{min} = 29.25$ 
& $\chi^2_{min} = 12.89$ \\
\cline{2-3}
& $x^* =(0.7757, 0.3211, 0.0000)$ 
& $x^* =(1.0690, 0.2938, 0.0000)$ \\
\hline\hline

\textbf{GW} 
& $\chi^2_{min} = 29.25$ 
& $\chi^2_{min} = 12.89$ \\
\cline{2-3}
& $x^* =(0.7757, 0.3211, 0.0000, 0.8461)$ 
& $x^* =(1.0690, 0.2938, 0.0000, 0.8779)$ \\
\hline\hline

\textbf{LKM} 
& $\chi^2_{min} = 22.67$ 
& $\chi^2_{min} = 11.83$ \\
\cline{2-3}
& $x^* =(0.2168, 0.3738, 0.1302)$ 
& $x^* =(0.2570, 0.3632, 0.1942)$ \\
\hline\hline

\textbf{LBM} 
& $\chi^2_{min} = 22.13$ 
& $\chi^2_{min} = 11.80$ \\
\cline{2-3}
& $x^* =(0.2019, 0.3571, 0.1016, 1.1415)$ 
& $x^* =(0.2499, 0.3547, 0.1832, 1.0588)$ \\
\hline\hline

\textbf{LBW} 
& $\chi^2_{min} = 20.51$ 
& $\chi^2_{min} = 11.57$ \\
\cline{2-3}
& $x^* =(0.1986, 0.3434, 0.0597, 1.2464, 0.5883)$ 
& $x^* =(0.2458, 0.3834, 0.2236, 0.9299, 1.0158)$ \\
\hline

$r-m-1$ 
& 26-5-1 = 20 
& 23-5-1 = 17 \\
\hline

$U_{0.95}$ 
& 31.41 (33.92 for $m=3$) 
& 27.58 (30.34 for $m=3$) \\
\hline

\end{tabular}
}

\noindent\textbf{Source:} Author's calculations based on data from the Human Mortality Database. 
Available at: \url{https://www.mortality.org/Country/Country?cntr=SWE}. 
Accessed: 2026-03-15.

\end{table}

Comments on the Table \ref{tab:sweden1900_80_120}:

(i) Just as in the case of the Norwegian population, the $\chi^2$ test accepted the $H(80, 120)$ hypothesis at the $p=0.05$ significance level for all distributions: $GM$, $GW$, $LKM$, $LBM$, $LBW$ for both Swedish women and men.

(ii) In our opinion, mortality data over the interval (80, 120) years for the population of Sweden, as in the case of Norway, are adequately described by any of the distributions presented in Section \ref{section2}.

(iii) However, it is impossible to assert, based on Tables \ref{tab:norway1900_80_120} and \ref{tab:sweden1900_80_120}, that the $GM$ distribution provided a better approximation to the actual data for Norway and Sweden over the (80, 120) year interval. For example, the $GM$ and $LKM$ distributions have the same number of parameters $m=3$, but in both cases, for both women and men, the $LKM$ distribution yielded a lower value of $\chi^2_{min}$ .

\vspace{0.5cm} 
{\it Example 6. England $\&$ Wales: cohort born in 1900, interval (80, 120) years.}

The following Table \ref{tab:england_wales1900_80_120} contains the results of calculations of the minimum values of $\chi^2$ statistics for mortality data separately for women and men in England $\&$ Wales over the interval (80, 120) years.

\begin{table}[h]
\centering
\small

\caption{England $\&$ Wales, cohort born in 1900, interval (80, 120) years}
\label{tab:england_wales1900_80_120}

\resizebox{\textwidth}{!}{
\begin{tabular}{|l|c|c|}
\hline
\textbf{Gender} & \textbf{Women} & \textbf{Men} \\
\hline\hline

$(t_0 , t_{r})$ 
& $(80 , 120)$ 
& $(80 , 120)$ \\
\hline

$N(t_0)$ 
& 154901 
& 76100 \\
\hline\hline

\textbf{GM} 
& $\chi^2_{min} = 36.81$ 
& $\chi^2_{min} = 33.48$ \\
\cline{2-3}
& $x^* =(0.8905, 0.2994, 0.0000)$ 
& $x^* =(1.1647, 0.2771, 0.1218)$ \\
\hline\hline

\textbf{GW} 
& $\chi^2_{min} = 35.26$ 
& $\chi^2_{min} = 24.96$ \\
\cline{2-3}
& $x^* =(0.8675, 0.3084, 0.0488, 1.1197)$ 
& $x^* =(0.8463, 0.3074, 0.2228, 1.0362)$ \\
\hline\hline

\textbf{LKM} 
& $\chi^2_{min} = 56.07$ 
& $\chi^2_{min} = 43.78$ \\
\cline{2-3}
& $x^* =(0.2167, 0.3612, 0.1660)$ 
& $x^* =(0.2650, 0.3435, 0.2336)$ \\
\hline\hline

\textbf{LBM} 
& $\chi^2_{min} = 35.08$ 
& $\chi^2_{min} = 41.74$ \\
\cline{2-3}
& $x^* =(0.1281, 0.3110, 0.0609, 2.0257)$ 
& $x^* =(0.2279, 0.2879, 0.1184, 1.4545)$ \\
\hline\hline

\textbf{LBW} 
& $\chi^2_{min} = 30.19$ 
& $\chi^2_{min} = 24.40$ \\
\cline{2-3}
& $x^* =(0.1710, 0.3452, 0.1553, 1.2774, 1.0357)$ 
& $x^* =(0.1470, 0.3340, 0.2515, 1.6007, 1.0332)$ \\
\hline

$r-m-1$ 
& 29-5-1 = 23 
& 27-5-1 = 21 \\
\hline

$U_{0.95}$ 
& 35.17 (37.65 for $m=3$) 
& 32.67 (35.17 for $m=3$) \\
\hline

\end{tabular}
}

\noindent\textbf{Source:} Author's calculations based on data from the Human Mortality Database. 
Available at: \url{https://www.mortality.org/Country/Country?cntr=GBRCENW}. 
Accessed: 2026-03-15.

\end{table}

Comments on the Table \ref{tab:england_wales1900_80_120}:

(i) The $\chi^2 $ test at the significance level of $p=0.05$ for women England $\&$ Wales accepted the $H(80, 120)$ hypothesis for the distributions: $GM$, $GW$, $LBM$, $LBW$ (sample size n=154901).

(ii) For England $\&$ Wales men, the $\chi^2$ criterion accepted this hypothesis for the distributions: $GM$, $GW$, $LBW$.

(iii) For the England $\&$ Wales population, the $LBM$ distribution was rejected (for men), and the $LKM$ distribution was similarly rejected (for both women and men).

(i$\vee)$ It should be noted that, for example, for women in Belgium, Portugal and Switzerland, born in 1900, the $\chi^2$ test at the $p=0.05$ significance level also accepted the $H(80, 120)$ hypothesis for the $LBW$ distribution (here we omit the corresponding calculations).

{\it Remark 4.} Note that data on mortality over the interval (80, 120) years for the population of England \& Wales, Norway and Sweden born in 1900 are most likely the best data of a sufficiently large volume that we analyzed. These data are close to an "ideal" sample in the context of applied statistics.

Unfortunately, in practice the situation is often not ''ideal''. For example, if, similarly to Table \ref{tab:sweden1900_80_120}, we analyze the mortality data for Swedish women born in 1881 over the interval (80, 120) years, then the hypothesis $H(80, 120)$ will be rejected by the $\chi^2 $-criterion at the significance level $p=0.05$ for two main distributions $GW$, $LBW$ (possibly for others).

In this case, the authors are not prepared to provide any analysis or explanation. All that remains is to once again refer to J. Neumann's words, quoted above.

{\it Conclusions.} A number of biologists and gerontologists believe that somewhere in the interval of (30,80) years (Gompertz period) the Gompertz-Makeham law often describes real data such as human life expectancy better than other distributions. And this is indeed true if we compare the $GM$ distribution with such well-known distributions as the normal, exponential and Weibull distributions.

However, the above calculation results based on real data for England and Wales, Norway, and Sweden generally do not support this hypothesis. Thus, from Tables \ref{tab:norway1900} - \ref{tab:engwales1900} (Section 4) and Table \ref{tab:engwales1900_thinned} (Section 5) it is clear that the Gompertz-Makeham distribution for the $\chi^2$ criterion on the interval (30,80) is inferior to such distributions as $GW$, $LBM$, $LBW$.

At the same time, the authors do not observe a significant difference between the distributions of $GM$ and $GW$, $LBW$ when used to analyze population mortality over the interval (80,120) years (see Tables \ref{tab:norway1900_80_120} - \ref{tab:england_wales1900_80_120} in Section 6).

Note that in our calculations the $LBW$ distribution based on the $\chi^2$ criterion generally showed the best results among all the distributions that were measured in this article, both in the interval (30, 80) years and in the final interval (80, 120) years.

It should be noted that the above conclusions are true only in the case of reliable data, for which the influence of unforeseen external factors does not significantly change the mortality rate.

\section*{Declarations}

\textbf{Funding} Not applicable. \\
\textbf{Conflict of interest/Competing interests}  The authors declare that they have no known competing financial interests or personal relationships that could have appeared to influence the work reported in this paper.\\
\textbf{Ethics approval and consent to participate} Not applicable. \\
\textbf{Consent for publication} Not applicable. \\
\textbf{Data availability} Not applicable. \\
\textbf{Materials availability} Not applicable. \\
\textbf{Code availability} Not applicable.  \\
\textbf{Author contribution}  All authors contributed equally to this article.

\clearpage

\section*{Appendix 1}
\begin{center}
\small % приблизно відповідає 10pt для таблиці
\begin{longtable}{r|cc|cc|cc}
\caption{Mortality data for the interval (30, 80+) years, cohort born in 1900} \\

\toprule
Age & \multicolumn{2}{c|}{Norway} & \multicolumn{2}{c|}{Sweden} & \multicolumn{2}{c}{England \& Wales} \\
 & Women & Men & Women & Men & Women & Men \\
\midrule
\endfirsthead

\toprule
Age & \multicolumn{2}{c|}{Norway} & \multicolumn{2}{c|}{Sweden} & \multicolumn{2}{c}{England \& Wales} \\
 & Women & Men & Women & Men & Women & Men \\
\midrule
\endhead

\bottomrule
\endfoot

\bottomrule
\endlastfoot

30 & 96 & 107 & 191 & 207 & 1109 & 1091 \\
31 & 94 & 92 & 164 & 196 & 1044 & 1079 \\
32 & 99 & 97 & 190 & 199 & 1125 & 1168 \\
33 & 86 & 98 & 186 & 174 & 1132 & 1179 \\
34 & 76 & 90 & 149 & 158 & 1116 & 1193 \\
35 & 80 & 86 & 152 & 143 & 1087 & 1198 \\
36 & 75 & 92 & 185 & 199 & 1158 & 1246 \\
37 & 73 & 104 & 180 & 176 & 1125 & 1263 \\
38 & 71 & 91 & 161 & 193 & 1069 & 1319 \\
39 & 71 & 97 & 156 & 200 & 1225 & 1478 \\
40 & 71 & 115 & 127 & 166 & 1364 & 1653 \\
41 & 73 & 120 & 158 & 181 & 1241 & 1568 \\
42 & 80 & 127 & 153 & 182 & 1284 & 1592 \\
43 & 90 & 129 & 171 & 189 & 1316 & 1585 \\
44 & 93 & 119 & 181 & 178 & 1296 & 1610 \\
45 & 82 & 106 & 202 & 219 & 1343 & 1724 \\
46 & 86 & 108 & 202 & 219 & 1431 & 1894 \\
47 & 88 & 111 & 201 & 224 & 1454 & 2024 \\
48 & 82 & 108 & 214 & 248 & 1515 & 2151 \\
49 & 81 & 112 & 215 & 272 & 1673 & 2386 \\
50 & 79 & 122 & 233 & 274 & 1744 & 2596 \\
51 & 82 & 132 & 236 & 283 & 1793 & 2808 \\
52 & 89 & 139 & 258 & 306 & 1921 & 3013 \\
53 & 106 & 146 & 292 & 319 & 2059 & 3304 \\
54 & 114 & 161 & 292 & 356 & 2161 & 3597 \\
55 & 111 & 169 & 288 & 449 & 2207 & 3872 \\
56 & 118 & 190 & 325 & 447 & 2452 & 4320 \\
57 & 127 & 201 & 340 & 514 & 2597 & 4762 \\
58 & 134 & 205 & 358 & 493 & 2819 & 5222 \\
59 & 155 & 236 & 361 & 573 & 2927 & 5393 \\
60 & 174 & 278 & 408 & 652 & 3233 & 5877 \\
61 & 184 & 317 & 417 & 642 & 3487 & 6386 \\
62 & 206 & 335 & 465 & 733 & 4031 & 7047 \\
63 & 222 & 372 & 534 & 840 & 4150 & 7340 \\
64 & 239 & 400 & 624 & 876 & 4353 & 7722 \\
65 & 258 & 426 & 553 & 892 & 4812 & 8072 \\
66 & 280 & 486 & 631 & 1014 & 5016 & 8271 \\
67 & 318 & 506 & 731 & 1111 & 5565 & 8874 \\
68 & 363 & 553 & 776 & 1204 & 6084 & 9313 \\
69 & 372 & 611 & 822 & 1213 & 6563 & 9788 \\
70 & 385 & 636 & 888 & 1263 & 6850 & 9680 \\
71 & 440 & 658 & 990 & 1471 & 7279 & 9912 \\
72 & 493 & 666 & 1065 & 1407 & 8165 & 10290 \\
73 & 529 & 685 & 1112 & 1523 & 8353 & 10412 \\
74 & 552 & 706 & 1176 & 1581 & 8968 & 10431 \\
75 & 588 & 691 & 1270 & 1647 & 9460 & 10236 \\
76 & 627 & 678 & 1346 & 1648 & 9804 & 10281 \\
77 & 681 & 706 & 1501 & 1703 & 10141 & 9678 \\
78 & 733 & 758 & 1503 & 1738 & 10481 & 9573 \\
79 & 734 & 750 & 1587 & 1684 & 10966 & 9189 \\
80+ & 12084 & 7434 & 26432 & 17220 & 154901 & 76100 \\

\end{longtable}
\noindent\textbf{Source:} Author's calculations based on data from the Human Mortality Database. 
Available at: \url{https://www.mortality.org/Data/DataAvailability}. 
Accessed: 2026-03-15.
\end{center}

\section*{Appendix 2}
\begin{center}
\small
\begin{longtable}{r|cc|cc|cc}
\caption{Mortality data for the interval (80, 120) years, cohort born in 1900} \\

\toprule
Age & \multicolumn{2}{c|}{Norway} & \multicolumn{2}{c|}{Sweden} & \multicolumn{2}{c}{England \& Wales} \\
 & Women & Men & Women & Men & Women & Men \\
\midrule
\endfirsthead

\toprule
Age & \multicolumn{2}{c|}{Norway} & \multicolumn{2}{c|}{Sweden} & \multicolumn{2}{c}{England \& Wales} \\
 & Women & Men & Women & Men & Women & Men \\
\midrule
\endhead

\bottomrule
\endfoot

\bottomrule
\endlastfoot

80 & 771 & 738 & 1686 & 1680 & 10978 & 8664 \\
81 & 769 & 685 & 1710 & 1656 & 11208 & 8319 \\
82 & 805 & 673 & 1749 & 1613 & 11411 & 7914 \\
83 & 830 & 648 & 1717 & 1516 & 11157 & 7169 \\
84 & 855 & 674 & 1809 & 1460 & 11211 & 6641 \\
85 & 806 & 557 & 1910 & 1340 & 11103 & 6151 \\
86 & 851 & 555 & 1855 & 1241 & 10522 & 5347 \\
87 & 809 & 545 & 1753 & 1170 & 9912 & 4595 \\
88 & 802 & 448 & 1745 & 1020 & 9498 & 4083 \\
89 & 813 & 405 & 1599 & 908 & 8852 & 3493 \\
90 & 683 & 312 & 1486 & 803 & 8036 & 2941 \\
91 & 645 & 280 & 1434 & 674 & 7294 & 2420 \\
92 & 523 & 230 & 1204 & 522 & 6633 & 2104 \\
93 & 530 & 215 & 1046 & 466 & 5832 & 1705 \\
94 & 413 & 140 & 858 & 314 & 4953 & 1287 \\
95 & 315 & 107 & 771 & 255 & 4141 & 1026 \\
96 & 238 & 80 & 616 & 206 & 3289 & 736 \\
97 & 200 & 53 & 454 & 136 & 2583 & 511 \\
98 & 140 & 31 & 349 & 85 & 1980 & 365 \\
99 & 105 & 22 & 243 & 61 & 1396 & 236 \\

100 & 63 & \ldelim\{{2}{3mm} \hspace{2mm}\multirow{2}{*}{19} & 158 & 34 & 1014 & 155 \\
101 & 53 &  & 120 & 31 & 690 & 100 \\

102 & 29 & \ldelim\{{8}{3mm} \hspace{2mm}\multirow{8}{*}{17} & 73 & 17 & 508 & 66 \\
103 & 16 &  & 36 & \ldelim\{{7}{3mm} \hspace{2mm}\multirow{7}{*}{12} & 308 & 37 \\
104 & \ldelim\{{6}{3mm} \hspace{2mm}\multirow{6}{*}{20} &  & 21 &  & 170 & 15 \\

105 &  &  & 14 &  & 119 & \ldelim\{{5}{3mm} \hspace{2mm}\multirow{5}{*}{21} \\

106 &  &  & \ldelim\{{4}{3mm} \hspace{2mm}\multirow{4}{*}{16} &  & 53 &  \\
107 &  &  &  &  & 28 &  \\
108 &  &  &  &  & \ldelim\{{2}{3mm} \hspace{2mm}\multirow{2}{*}{24} &  \\
109+ &  &  &  &  &  &  \\

\end{longtable}
\end{center}

\noindent\textbf{Source:} Author's calculations based on data from the Human Mortality Database. 
Available at: \url{https://www.mortality.org/Data/DataAvailability}. 
Accessed: 2026-03-15.

%\textbf{Competing Interests.} The authors declare that they have no competing interests.

\vspace{0.5cm}

%\newpage

\end{document}